\documentclass[aps,prl,showpacs,twocolumn,a4]{revtex4}
\usepackage{graphicx}
\usepackage{graphicx}  
\usepackage{dcolumn}   
\usepackage{bm}        
\usepackage{amssymb}   
\begin{document}
\onecolumngrid
\begin{center}
\bf \begin{boldmath} Measurement 
of the Cross Section for  
Prompt Diphoton Production \\ in $p\overline{p}$ 
Collisions at $\sqrt{s} = 1.96 ~\hbox{TeV}$ 
 \end{boldmath}
\end{center}




\date{\today}
\thispagestyle{empty}
\hfilneg
\font\eightit=cmti8
\def\r#1{\ignorespaces $^{#1}$}
\hfilneg
\begin{sloppypar}
\noindent 
D.~Acosta,\r {16} J.~Adelman,\r {12} T.~Affolder,\r 9 T.~Akimoto,\r {54}
M.G.~Albrow,\r {15} D.~Ambrose,\r {43} S.~Amerio,\r {42}  
D.~Amidei,\r {33} A.~Anastassov,\r {50} K.~Anikeev,\r {15}
% ...
\end{sloppypar}
\vskip .026in
\begin{center}
(CDF Collaboration)
\end{center}

\vskip .026in
\begin{center}
\r 1  {\eightit Institute of Physics, Academia Sinica, Taipei, Taiwan 11529, 
Republic of China} \\
\r 2  {\eightit Argonne National Laboratory, Argonne, Illinois 60439} \\
\r 3  {\eightit Institut de Fisica d'Altes Energies, Universitat Autonoma
de Barcelona, E-08193, Bellaterra (Barcelona), Spain} \\
\r 4  {\eightit Istituto Nazionale di Fisica Nucleare, University of Bologna,
I-40127 Bologna, Italy} \\
\r 5  {\eightit Brandeis University, Waltham, Massachusetts 02254} \\
\r 6  {\eightit University of California at Davis, Davis, California  95616} \\
\r 7  {\eightit University of California at Los Angeles, Los 
Angeles, California  90024} \\
\r 8  {\eightit University of California at San Diego, La Jolla, California  92093} \\ 
\r 9  {\eightit University of California at Santa Barbara, Santa Barbara, California 
93106} \\ 
\r {10} {\eightit Instituto de Fisica de Cantabria, CSIC-University of Cantabria, 
39005 Santander, Spain} \\
\r {11} {\eightit Carnegie Mellon University, Pittsburgh, PA  15213} \\
\r {12} {\eightit Enrico Fermi Institute, University of Chicago, Chicago, 
Illinois 60637} \\
\r {13}  {\eightit Joint Institute for Nuclear Research, RU-141980 Dubna, Russia}
\\
\r {14} {\eightit Duke University, Durham, North Carolina  27708} \\
\r {15} {\eightit Fermi National Accelerator Laboratory, Batavia, Illinois 
60510} \\
\r {16} {\eightit University of Florida, Gainesville, Florida  32611} \\
\r {17} {\eightit Laboratori Nazionali di Frascati, Istituto Nazionale di Fisica
               Nucleare, I-00044 Frascati, Italy} \\
\r {18} {\eightit University of Geneva, CH-1211 Geneva 4, Switzerland} \\
\r {19} {\eightit Glasgow University, Glasgow G12 8QQ, United Kingdom}\\
\r {20} {\eightit Harvard University, Cambridge, Massachusetts 02138} \\
\r {21} {\eightit The Helsinki Group: Helsinki Institute of Physics; and Division of
High Energy Physics, Department of Physical Sciences, University of Helsinki, FIN-00044, Helsinki, Finland}\\
\r {22} {\eightit Hiroshima University, Higashi-Hiroshima 724, Japan} \\
\r {23} {\eightit University of Illinois, Urbana, Illinois 61801} \\
\r {24} {\eightit The Johns Hopkins University, Baltimore, Maryland 21218} \\
\r {25} {\eightit Institut f\"{u}r Experimentelle Kernphysik, 
Universit\"{a}t Karlsruhe, 76128 Karlsruhe, Germany} \\
\r {26} {\eightit High Energy Accelerator Research Organization (KEK), Tsukuba, 
Ibaraki 305, Japan} \\
\r {27} {\eightit Center for High Energy Physics: Kyungpook National
University, Taegu 702-701; Seoul National University, Seoul 151-742; and
SungKyunKwan University, Suwon 440-746; Korea} \\
\r {28} {\eightit Ernest Orlando Lawrence Berkeley National Laboratory, 
Berkeley, California 94720} \\
\r {29} {\eightit University of Liverpool, Liverpool L69 7ZE, United Kingdom} \\
\r {30} {\eightit University College London, London WC1E 6BT, United Kingdom} \\
\r {31} {\eightit Massachusetts Institute of Technology, Cambridge,
Massachusetts  02139} \\   
\r {32} {\eightit Institute of Particle Physics: McGill University,
Montr\'{e}al, Canada H3A~2T8; and University of Toronto, Toronto, Canada
M5S~1A7} \\
\r {33} {\eightit University of Michigan, Ann Arbor, Michigan 48109} \\
\r {34} {\eightit Michigan State University, East Lansing, Michigan  48824} \\
\r {35} {\eightit Institution for Theoretical and Experimental Physics, ITEP,
Moscow 117259, Russia} \\
\r {36} {\eightit University of New Mexico, Albuquerque, New Mexico 87131} \\
\r {37} {\eightit Northwestern University, Evanston, Illinois  60208} \\
\r {38} {\eightit The Ohio State University, Columbus, Ohio  43210} \\  
\r {39} {\eightit Okayama University, Okayama 700-8530, Japan}\\  
\r {40} {\eightit Osaka City University, Osaka 588, Japan} \\
\r {41} {\eightit University of Oxford, Oxford OX1 3RH, United Kingdom} \\
\r {42} {\eightit University of Padova, Istituto Nazionale di Fisica 
          Nucleare, Sezione di Padova-Trento, I-35131 Padova, Italy} \\
\r {43} {\eightit University of Pennsylvania, Philadelphia, 
        Pennsylvania 19104} \\   
\r {44} {\eightit Istituto Nazionale di Fisica Nucleare, University and Scuola
               Normale Superiore of Pisa, I-56100 Pisa, Italy} \\
\r {45} {\eightit University of Pittsburgh, Pittsburgh, Pennsylvania 15260} \\
\r {46} {\eightit Purdue University, West Lafayette, Indiana 47907} \\
\r {47} {\eightit University of Rochester, Rochester, New York 14627} \\
\r {48} {\eightit The Rockefeller University, New York, New York 10021} \\
\r {49} {\eightit Istituto Nazionale di Fisica Nucleare, Sezione di Roma 1,
University di Roma ``La Sapienza," I-00185 Roma, Italy}\\
\r {50} {\eightit Rutgers University, Piscataway, New Jersey 08855} \\
\r {51} {\eightit Texas A\&M University, College Station, Texas 77843} \\
\r {52} {\eightit Texas Tech University, Lubbock, Texas 79409} \\
\r {53} {\eightit Istituto Nazionale di Fisica Nucleare, University of Trieste/\
Udine, Italy} \\
\r {54} {\eightit University of Tsukuba, Tsukuba, Ibaraki 305, Japan} \\
\r {55} {\eightit Tufts University, Medford, Massachusetts 02155} \\
\r {56} {\eightit Waseda University, Tokyo 169, Japan} \\
\r {57} {\eightit Wayne State University, Detroit, Michigan  48201} \\
\r {58} {\eightit University of Wisconsin, Madison, Wisconsin 53706} \\
\r {59} {\eightit Yale University, New Haven, Connecticut 06520} \\
\end{center}

\begin{abstract}
We report a measurement of the rate of prompt diphoton production 
in $p\overline{p}$ 
collisions at $\sqrt{s}=1.96 ~\hbox{TeV}$ using a data sample of 
 207 pb$^{-1}$ collected with the upgraded Collider Detector at 
Fermilab (CDF II). The background from non-prompt sources is 
determined using a statistical method based on differences in the
 electromagnetic 
 showers. The cross section is measured as a function of 
the diphoton mass, the transverse momentum of the
 diphoton system, and the azimuthal angle between the 
two photons and is found to be consistent
 with perturbative QCD predictions. 
\end{abstract} 


\maketitle
\twocolumngrid
Diphoton ($\gamma\gamma$) final states are a signature
 of many interesting physics processes. 
For example, at the LHC, one  of the main discovery channels 
 for the Higgs search is the $\gamma\gamma$ final state~\cite{ATLAS, CMS}. 
 An excess of $\gamma\gamma$ production at high invariant mass could be
 a signature of large extra dimensions~\cite{extradim},
 and in many theories involving physics beyond the standard model, 
cascade decays of heavy new particles generate a $\gamma\gamma$ 
signature~\cite{newphysics}. However, the QCD production rate is 
large compared to most new physics, so an understanding 
of the QCD production mechanism is a prerequisite to searching
 reliably for new physics in this channel. In addition, the  two-photon 
final state is interesting in its own right. 
Due to the excellent energy resolution of the CDF electromagnetic (EM) 
 calorimeter, 
the 4-momenta of the two photons in the final state 
can be determined with good precision. This allows, for example, 
a direct measurement of the transverse momentum of 
the $\gamma\gamma$ system ($q_{T}$),
 which is sensitive to initial state soft gluon radiation.

In perturbative Quantum Chromodynamics (pQCD),
 the leading contributions are from 
 quark anti-quark annihilation ($q\overline{q}\rightarrow\gamma\gamma$)  
and gluon-gluon scattering ($gg\rightarrow\gamma\gamma$). 
The latter subprocess involves initial state gluons coupling to 
 the  final  state photons through 
 a quark box; thus, this  subprocess is suppressed by
 a factor of $\alpha_s^2$ with respect to the $q\overline{q}$ subprocess. 
However,  the rate is still appreciable in kinematic
 regions where the $gg$ parton luminosity is high, 
such as at low $\gamma\gamma$ mass. Because the probability for
 a hard parton to fragment to a photon 
is of order $\alpha_{em}/\alpha_{s}$, 
processes involving  the production of one (zero) prompt
 photons and one (two) photons originating from 
parton fragmentation are also effectively of leading order (LO).
 Next-to-leading order (NLO) contributions 
include real and virtual corrections to the above subprocesses.

 We have compared our experimental results to 
three predictions : DIPHOX~\cite{DIPHOX}, ResBos~\cite{ResBos}, 
and PYTHIA~\cite{pythia}.
 DIPHOX is a fixed-order QCD calculation that
 includes all of the above subprocesses at 
 NLO (except for $gg\rightarrow\gamma\gamma$ which
 is present only at LO). Recently, NLO corrections for
$gg\rightarrow\gamma\gamma$  have been calculated~\cite{bern} and
 we have added these corrections to the DIPHOX prediction.
 The ResBos program includes  
 subprocesses where the two photons are 
produced at the hard-scattering at NLO and 
fragmentation contributions at LO; but also 
resums the effects of initial state soft gluon
 radiation. This  is particularly important for 
examination of the $\gamma\gamma$ $q_T$ 
distribution, which is a delta function at LO
 and divergent as $q_T\rightarrow0$ at NLO, 
and thus requires a soft gluon resummation in
 order to provide a physical description of 
the $\gamma\gamma$ data in this region. PYTHIA 
is a parton shower Monte Carlo program that 
contains the above processes at LO. 

At hadron-hadron colliders, it is difficult to measure
 a fully inclusive $\gamma\gamma$ cross section due to
 the large backgrounds 
from quarks and gluons fragmenting into neutral mesons 
which carry most of the parent parton's momentum.  
Isolation requirements are typically used to reduce 
these backgrounds.
In this analysis, the isolation criterion requires 
that the transverse energy ($E_T$) sum in a
 cone of radius  $R=0.4$(in $\eta-\phi$ space)~\cite{CDFAxis}
 about the photon direction, minus the photon energy, be less than 1 GeV. 
This isolation requirement reduces the backgrounds from neutral
 mesons decaying into photons and photon production from fragmentation sources.
 The CDF isolation requirement effectively removes all
 contributions where both photons originate from fragmentation subprocesses.
 However, as will be noted later, some indication of single
 fragmentation subprocesses can still be observed  in the CDF data. 

The CDF II detector is a magnetic spectrometer which is described 
in detail elsewhere~\cite{CDFIITDR}. 
The central detector consists of a silicon micro-strip vertex detector 
inside a cylindrical drift chamber, both of which are immersed
 in the 1.4 T 
magnetic field of a superconducting solenoid.  
Outside the solenoid is the central calorimeter 
which is divided 
into an electromagnetic 
compartment (CEM) on the inside and hadronic compartment 
(CHA) on the outside.
  Both calorimeters are segmented into towers of granularity 
$\Delta \eta \times \Delta\phi \approx  0.1 \times 0.26$. 
The CEM consists  of a  scintillator-lead
 calorimeter along with an embedded multi-wire proportional 
chamber (CES) located near shower maximum at 6 radiation lengths. 
The CES  allows for a position determination of the EM shower and 
for a  measurement of the lateral shower profile. 
The average energy resolution of the CEM is 
$\sigma(E) / E = 13.5\%/\sqrt{E \sin \theta}$ (with E in GeV)
 and the position resolution of the CES is 2 mm  for a 50 GeV photon.
 Another important component for this analysis is a preshower wire chamber
 (CPR) mounted between the magnet coil and the CEM, 
at about 1.2/$\sin\theta$ radiation lengths. 
The CPR detects photon candidates that have converted in
 the magnet coil and other material in the inner detector. 

This analysis~\cite{phd} uses events collected with a 
trigger that requires two photon candidates with $E_{T}$ greater 
than 12 GeV each. A requirement of $E_T$ greater than 14 GeV (13 GeV) 
for the leading (softer)
photon candidate in the event is imposed in the offline analysis.
 The minimum transverse energy requirements for the two photon
 candidates are different in order to avoid the kinematic region 
where the NLO calculation is unstable due to the imperfect
 cancellation of the real and virtual gluon divergences.

In identifying photons, 
we impose fiducial requirements to avoid uninstrumented regions at 
the edges of the CES; as part of this criterion we require
 the pseudorapidity of the photon candidate to be in the 
interval $|\eta|<0.9$. The reconstructed z-vertex for the collision
 is required to be less than 60 cm from the center of the detector. 
 The ratio of the  hadronic energy to EM energy (Had/EM) for the 
photon candidates must be less than 0.055+0.00045$\times E$,
 with $E$ the EM energy in GeV.  The isolation energy 
is required to be below 1 GeV. 
Although only about 1\% of showers from prompt photons have more
 than 1 GeV of additional energy in the isolation cone, about 
15\% of the photon showers fail the isolation requirement because 
of additional energy from the underlying event~\cite{isolation}. 
Photon candidates with any tracks ($p_T$ above 0.5 GeV)
 that can be extrapolated to them are rejected.
 The lateral profile of EM showers in the CES is compared
 to the profile of electrons measured in a test beam. 
The definition of the $\chi^2$ from the comparison can be found 
in Ref.~\cite{Run0_phoPRD}. 
We require the $\chi^2$ of the comparison to be less than 20
 in the event selection and reject photon candidates with an 
additional CES cluster above 1 GeV~\cite{photon_prd}.
The efficiencies for each event selection requirement,
 evaluated using a combination of PYTHIA Monte Carlo and data,
 are listed sequentially in Table \ref{eff_table}. 
 The trigger efficiency per photon, measured
using a single photon trigger, is approximately 80\% 
at 13 GeV and rises
to greater than 99\% for photons above 15 GeV. 
The combined selection efficiency ($\varepsilon_{tot}$), 
including acceptance and trigger efficiency, is 15.2\% per diphoton event.  

\begin{table} 
\caption {The selection efficiencies per diphoton event.} 
\begin{center}
\begin{tabular}{ll}
\hline
\hline 
Trigger efficiency & 0.951 \\
Reconstruction efficiency and fiducial & 0.423  \\ 
Isolation energy in 0.4 cone $<$ 1 GeV & 0.727 \\
No track pointing to the EM cluster & 0.699\\ 
No extra CES cluster above 1 GeV & 0.899 \\ 
CES $\chi^{2}$ $<$ 20 & 0.970 \\ 
Had/EM $<$0.055 + 0.00045$\times E$  &  0.976 \\ 
$|$z-vertex $|<$ 60 cm & 0.877 \\ 
\hline 
\hline  
Combined ($\varepsilon_{tot}$) & 0.152 \\ 
\end{tabular} 
\end{center}
\label{eff_table} 
\end{table}

\begin{figure}
\begin{center}
\resizebox{8cm}{5cm}{\includegraphics{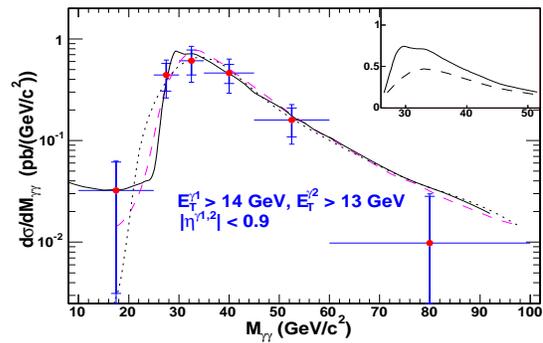}}
\end{center}
\caption{The $\gamma\gamma$ mass distribution from the
 CDF Run II data, along with predictions from  DIPHOX 
(solid), ResBos (dashed), and PYTHIA (dotted).
 The PYTHIA predictions have been scaled by a factor
 of 2. The inset shows, on a linear scale,
 the total $\gamma\gamma$ cross section in DIPHOX
 with (solid)/without (dashed) the $gg$ contribution. }

\label{mass_205}
\end{figure}

\begin{figure}
\begin{center}
\resizebox{8cm}{5cm}{\includegraphics{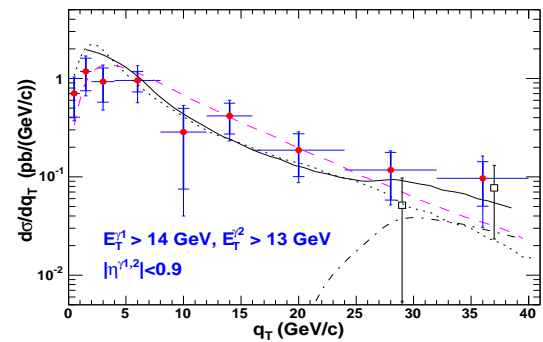}}
\end{center}
\caption{The $\gamma\gamma$  $q_{T}$ distribution 
from the CDF Run II data, along with predictions from DIPHOX (solid),
 ResBos (dashed), and PYTHIA (dotted).
 The PYTHIA predictions have been scaled by a factor of 2.
  Also shown,  at larger $q_T$, are the DIPHOX prediction (dot-dashed)
 and the CDF Run 
II data (open squares : shifted to the right by 1 GeV for visibility)
 for the configuration where the two photons are required 
to have $\Delta\phi < \pi/2$. }
\label{qt_205}

\end{figure}

\begin{figure}
\begin{center}
\resizebox{8cm}{5cm}{\includegraphics{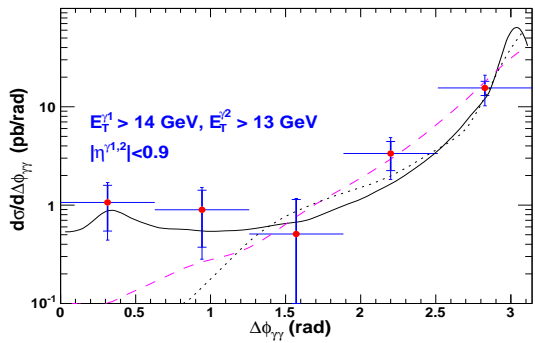}}
\end{center}
\caption{The $\Delta\phi$ angle  between the two 
photons from the CDF Run II data, along with predictions 
from DIPHOX (solid), ResBos (dashed), and PYTHIA (dotted).
 The PYTHIA predictions have been scaled by a factor of 2. 
The bump at $\Delta\phi = 0.3$ on the DIPHOX curve is unphysical; 
it is caused by the combination of collinear divergence and 
the anti-collinear cut ~\cite{phd}. }

\label{dphi_207}
\end{figure}

After imposing all of the requirements, 
889 two-photon candidates remain in our data sample.
 This sample includes  background from neutral mesons
 such as $\pi^{0}$ and $\eta$ that decay to multiple photons.
 To estimate this background, we apply the statistical background 
subtraction method described in~\cite{RunIADIPHO}, which makes use
 of the differences on average between EM showers produced by single 
photons and by the multiple photons produced in neutral  meson decays. 
The separation between single and multiple photon showers
 relies on the shower shape measured by the CES $\chi^2$ and 
the preshower
 conversion pulse height measured by the CPR. 
Since photons from the decay of neutral mesons
 with $E_{T}$ above 35 GeV 
are almost collinear in the lab frame, 
 their shower shape in the CES is no 
longer distinguishable from a single-photon shower. 
To estimate the background contamination in this 
high $E_{T}$ region, the CPR has been utilized. 
 The chance for a conversion to take place in the 
tracking volume or magnet coil (1.1 radiation lengths)
 and generating a hit in the CPR is higher for the multiple
 photons than for a single photon.
 We use the CES shower shape for photon showers with
  $E_T < $ 35 GeV and the CPR pulse height for $E_T > $ 35 GeV.
 For each photon candidate, we  test whether the CES $\chi^2 $  is 
less than 4 (low $E_T$) or the photon canidate produces a 
pulse height in the CPR greater than one minimum ionizing 
particle (high $E_T$).  
There are four possibilities for the final state:
 both candidates pass the test, 
the first candidate passes and the second fails,
 the first fails and the second passes, or both candidates 
 fail (the first candidate has the higher $E_T$).
 From the known efficiencies for photons and
 background to pass the $\chi^2$ and conversion tests,
 we can then determine the number of true $\gamma\gamma$ 
events (as well as the number of $\gamma$-background, 
background-$\gamma$ and background-background events).
 Using the two background techniques discussed,
 we determine that of the 889 candidates,  $427\pm59(stat)$  
are real $\gamma\gamma$ events. From these events, 
the calculated acceptance and the integrated luminosity, 
we determine the diphoton cross sections for several kinematic variables.
The $\gamma\gamma$ mass distribution is shown in Fig.~\ref{mass_205},
 along with predictions from  DIPHOX, ResBos and PYTHIA. 
 The $q_{T}$ distribution is shown in Fig. \ref{qt_205}, 
and the $\Delta\phi$ distribution between the two photons is shown in 
Fig.~\ref{dphi_207}. The vertical error bars on the data indicate 
the combined statistical and systematic uncertainties with the inner
 tick marks indicating the statistical uncertainty alone~\cite{hepdata}. 
 The PYTHIA predictions have been scaled (factor of 2) to 
the total measured cross section in all the figures. 
The cross sections as a function of these three different variables are also 
tabulated in Tables \ref{xsec_vs_mass_table}, \ref{xsec_vs_qT_table} and \ref{xsec_vs_dphi_table}.

 It should be noted that the background to the $\gamma\gamma$ signal
 has been determined independently for each kinematic bin as the background 
fraction can vary with the kinematics.
 Determining the background on a bin-by-bin basis
 increases the statistical uncertainty but decreases 
the systematic uncertainty.

The systematic effects include uncertainties on
 the selection efficiencies (11\%), uncertainties 
from the background subtraction (20-30\%)  and from
 the luminosity determination (6\%)~\cite{lum}.

\begin{table}
\caption{A comparison of the cross section as a function of the 
 $\gamma\gamma$ mass  for the data and predictions from  DIPHOX, ResBos and PYTHIA. }
\begin{tabular}{ccccc}
\hline
\hline
$M_{\gamma\gamma}$ &  CDF Data  & DIPHOX & ResBos & PYTHIA \\
 (GeV) & (pb/GeV)  & (pb/GeV) & (pb/GeV) & (pb/GeV) \\
\hline
 10-25 & 0.03 $\pm$ 0.03 $\pm$ 0.01 & 0.04 & 0.01 & 0.01\\
 25-30 & 0.44 $\pm$ 0.13 $\pm$ 0.12 & 0.41 & 0.31 & 0.18\\
 30-35 & 0.61 $\pm$ 0.17 $\pm$ 0.16 & 0.70 & 0.65 &  0.38\\
 35-45  & 0.46 $\pm$ 0.10 $\pm$ 0.14 & 0.46 & 0.43 & 0.24\\
 45-60  & 0.16 $\pm$ 0.05 $\pm$ 0.04 & 0.19 & 0.16 & 0.09\\
 60-100 & 0.01 $\pm$ 0.02 $\pm$ 0.01 & 0.04 & 0.04 & 0.02\\
\hline
\hline
\end{tabular}

\label{xsec_vs_mass_table}
\end{table}
\begin{table}
\caption{A comparison of the cross section as a function of the $\gamma\gamma$ $q_T$  for the data and predictions from  DIPHOX, ResBos and PYTHIA.}
\begin{tabular}{ccccc}
\hline
\hline
 $q_T$  & CDF Data & DIPHOX & ResBos & PYTHIA\\
 (GeV) &  (pb/GeV) & (pb/GeV) & (pb/GeV) & (pb/GeV) \\
\hline
 0-1 &  0.70 $\pm$ 0.30 $\pm$ 0.14 & -2.45  & 0.34 & 0.53\\
1-2 & 1.18 $\pm$ 0.43 $\pm$ 0.28 & 5.59 & 0.95 & 1.15\\
2-4 &  0.92 $\pm$ 0.35 $\pm$ 0.28  & 2.06 & 1.03 &  0.94\\
4-8  & 0.96 $\pm$ 0.23 $\pm$ 0.32 & 1.17 & 0.94 & 0.46\\
8-12 & 0.29 $\pm$ 0.21 $\pm$ 0.13 & 0.44 & 0.59 & 0.21\\
12-16 & 0.42 $\pm$ 0.14  $\pm$ 0.12 & 0.24 & 0.36 & 0.12\\
16-24 &  0.19 $\pm$ 0.09 $\pm$ 0.05 & 0.13 & 0.19 & 0.07\\
24-32 &  0.12 $\pm$ 0.06 $\pm$ 0.03 & 0.09 & 0.07 & 0.03\\
32-40 &  0.10 $\pm$ 0.05 $\pm$ 0.05 & 0.06 & 0.03 & 0.01\\
\hline
\hline
\end{tabular}

\label{xsec_vs_qT_table}
\end{table}

\begin{table}
\caption{A comparison of the cross section as a function of the $\gamma\gamma$ $\Delta\phi$  for the data and predictions from DIPHOX, ResBos and PYTHIA.}

\begin{tabular}{ccccc}
\hline
\hline
$\Delta\phi_{\gamma\gamma}$ & CDF Data & DIPHOX & ResBos & PYTHIA\\
($\pi$ rad) & (pb/rad)  & (pb/rad) & (pb/rad) & (pb/rad) \\
\hline
0.0-0.2 & 1.06 $\pm$ 0.52 $\pm$ 0.34 & 0.69 &  0.01 & 0.02\\
0.2-0.4 & 0.89 $\pm$ 0.52 $\pm$ 0.32 &   0.56 &  0.23 & 0.09\\
0.4-0.6 & 0.51 $\pm$ 0.63 $\pm$ 0.19 &0.71 & 0.73 &  0.44\\
0.6-0.8 & 3.34 $\pm$ 1.10 $\pm$ 1.04 & 1.83 & 3.08 & 1.09\\
0.8-1.0 & 15.56 $\pm$ 2.59 $\pm$ 4.70 & 23.37 & 17.52 &10.68\\
\hline
\hline
\end{tabular}

\label{xsec_vs_dphi_table}
\end{table}

We note some features of the theoretical predictions. 
The $gg$ subprocess provides a significant contribution to diphoton 
production at low mass ($\sim$30 GeV/$c^2$); the slight wiggle observed 
in the 
DIPHOX prediction is due to the 
very rapid falloff (with diphoton mass) 
of the $gg$ subprocess compared to the $q\overline{q}$ subprocess.
The ResBos $q_T$ curve is smooth for the entire range, 
while the  DIPHOX curve (as can be seen by the negative value in the first bin of Table~\ref{xsec_vs_qT_table}) is unstable at low $q_{T}$ due
 to the singularity noted earlier,  and thus is not plotted in Fig. ~\ref{qt_205} for $q_T < 2$ GeV/{\it c} ~\cite{Sudakov}.
At the high $q_T$ end, 
DIPHOX displays a shoulder, 
a feature absent in the ResBos prediction. 
The ResBos curve lies above the DIPHOX one
 at $\Delta\phi$ values of the order of $\pi/2$ but
 also lies significantly below the DIPHOX curve at small $\Delta\phi$.

The observed differences between the predictions are expected. 
 The fragmentation contribution in ResBos is effectively at LO. Since 
 fragmentation to a photon is  of order $\alpha_{em}/\alpha_{s}$,
 some 2$\rightarrow$3 processes such as $qg\rightarrow gq\gamma$,
 where the quark in the final state fragments to a second photon, are 
of order $\alpha_{em}^{2}\alpha_{s}$  and are included in a full NLO
 calculation. These contributions are present in DIPHOX, but not in ResBos,
 which leads to an underestimate of the production rate in the latter 
at high $q_{T}$, low $\Delta\phi$, and low $\gamma\gamma$ mass.  
In particular, the shoulder at $q_T$ of approximately 30 GeV/{\it c}
 arises from an increase in phase space for both the direct and fragmentation
 subprocesses~\cite{DIPHOXFrag}. It is instructive to divide the DIPHOX 
predictions into two regions : $\Delta\phi > \pi/2$ and $\Delta\phi < \pi/2$.
 We do so, and plot the $q_T$ prediction for the  
$\Delta\phi < \pi/2$ region in Fig.~\ref{qt_205} in order to highlight this
 contribution. It is apparent that the bump in the DIPHOX prediction at
 a $q_T$ of approximately 30 GeV/{\it c} is due  to the ``turn-on'' 
of the $\Delta\phi<\pi/2$ region of phase space. At $\Delta\phi$ values
 above $\pi/2$, the effects from soft gluon emission (included in ResBos
 but not in DIPHOX) are significant.

The data are in good agreement with the predictions for the mass distribution. 
At low to moderate $q_{T}$ and $\Delta\phi$ 
 greater than $\pi/2$, where the effect of soft 
gluon emissions are important, the data agree better with ResBos than DIPHOX.
  By contrast, in the regions where the 2$\rightarrow$3  
fragmentation contribution becomes important, large $q_{T}$,  $\Delta\phi$ 
 less than $\pi/2$ and low diphoton mass, the data agree better with DIPHOX. 


In this paper, we have presented results 
for $\gamma\gamma$ production in $p\overline{p}$ 
collisions at a center-of-mass energy of 1.96 TeV using 
a data sample twice that previously available. 
Good agreement has been observed with resummed and
 NLO predictions in different regions of phase space.
 For agreement in all areas, however, a resummed full NLO
 calculation will be necessary. 

We thank the Fermilab staff and the
 technical staffs of the participating institutions
 for their vital contributions.
 This work was supported by the U.S. Department 
 of Energy and National Science Foundation;
 the Italian Istituto Nazionale di Fisica Nucleare; 
the Ministry of Education, Culture, Sports, Science
 and Technology of Japan; the Natural Sciences and 
Engineering Research Council of Canada; the National 
Science Council of the Republic of China; the Swiss
 National Science Foundation; the A.P. Sloan Foundation; 
the Bundesministerium fuer Bildung und Forschung, Germany; 
the Korean Science and Engineering Foundation and the Korean 
Research Foundation; the Particle Physics and Astronomy Research
 Council and the Royal Society, UK; the Russian Foundation for
 Basic Research; the Comision Interministerial de Ciencia y
 Tecnologia, Spain; and in part by the European Community's 
Human Potential Programme under contract HPRN-CT-2002-00292, 
Probe for New Physics.

We would like to thank C.~Balazs,
  J.~Ph.~Guillet, E.~Pilon, C.~Schmidt 
and C.-P.~Yuan for invaluable discussions.


\begin{thebibliography}{999}

\bibitem{ATLAS} ATLAS 
Collaboration, {\it Technical Proposal},
 LHCC/P2 (1994); ATLAS Collaboration,
 {\it Physics Technical Design Report}, LHCC/99-15.
\bibitem{CMS} CMS Collaboration, {\it Technical Proposal}, LHCC/P1 (1994). 
\bibitem{extradim} B. Abbott {\it et al.}, 
Phys.\ Rev.\ Lett.\ {\bf 86}, 1156 (2001). 
\bibitem{newphysics} See, 
for example, G.~F.~Giudice and
 R.~Rattazzi, Phys.\ Rep.\ {\bf 322}, 41 (1999) 
and references therein. 
\bibitem{DIPHOX} T.~Binoth, J.~Ph.~Guillet, E.~Pilon and M.~Werlen, 
Eur.\ Phys.\ J.\ C \ {\bf 16}, 311 (2000).

\bibitem{ResBos} C.~Balazs, E.~L.~ Berger, S.~Mrenna and C.-P.~Yuan.
Phys.\ Rev.\ D\ {\bf 57}, 6934 (1998).
\bibitem{pythia} T. Sjostrand, P. Eden, C. Friberg, 
L. Lonnblad, G. Miu, S. Mrenna and     E. Norrbin,
Computer Physics Commun. {\bf 135 }(2001) 238.
The PYTHIA version used in this analysis is 6.216.

\bibitem{bern} Z.~Bern, L.~J.~Dixon and C.~Schmidt,
Nucl.\ Phys.\ Proc.\ Suppl.\  {\bf 116}, 178 (2003).
[arXiv:hep-ph/0211216].




\bibitem{CDFAxis}In  the CDF coordinate system, 
$\theta$ and $\phi$ are the polar and azimuthal angles, 
respectively, defined with respect to the proton beam direction, z.
 The pseudorapidity $\eta$ is defined as $-\ln(\tan(\theta/2))$.
 The transverse energy of a particle is $E_T=E \sin(\theta)$.
\bibitem{CDFIITDR} {\it The CDF II Detector Technical Design Report}, 
 CDF collaboration, R.~Blair et al., 
FERMILAB-Pub-96/390-E 

\bibitem{phd} Y.~Liu, FERMILAB-THESIS-2004-37. 
\bibitem{isolation}The photon isolation requirement efficiency 
is slightly $E_T$-dependent : 85$\%$ at 25 GeV, 78$\%$ at 60 GeV. 
Events in which a higher $E_T$ photon is produced also tend to
 create a greater amount of ambient energy in the calorimeter, 
due to the effects of initial state soft  gluon radiation. 
\bibitem{Run0_phoPRD}
F.~Abe {\it et al.}
Phys.\ Rev.\ D {\bf 48}, 2998 (1993).
\bibitem{photon_prd}
D.~Acosta {\it et al.},
Phys.\ Rev.\ D {\bf 65}, 112003 (2002)
[arXiv:hep-ex/0201004].
\bibitem{RunIADIPHO} F. Abe {\it et al.},
Phys. Rev. Lett. {\bf 70}, 2232 (1993). 
\bibitem{hepdata} Cross section
 tables for the different diphoton observables can be found
 at the HEPDATA database, durpdg.dur.ac.uk/HEPDATA/;
 see also [arXiv:hep-ex/0412050].
\bibitem{lum} S. Klimenko, J. Konigsberg and T.~M. Liss, Fermilab-FN-0741 (unpublished). 
\bibitem{Sudakov}In addition, a double
 logarithmic divergence is present
 when the $\gamma\gamma$ $q_T$ is exactly 
 equal to the maximum hadronic energy allowed 
 in the isolation cone. This is an example of what
 is known as a Sudakov shoulder. Since the logarithmic 
singularity is  integrable over the bin size, no divergence 
is actually produced, but the calculational instability remains. 
The divergence is not present in the ResBos prediction, 
since it  only appears at NLO in the fragmentation contribution.
 In the DIPHOX calculation, we have required less than 4 GeV of
 additional energy in the isolation cone, compared to the 1 GeV 
used in the experimental analysis. This looser theoretical isolation
 requirement improves the stability of the theory without having a
 significant impact on the numerical prediction.
\bibitem{DIPHOXFrag} T.~Binoth, J.~Ph.~Guillet, E.~Pilon and M.~Werlen, 
Phys.\ Rev.\ D\ {\bf 63}, 114016 (2003).

\end{thebibliography}
\end{document}